\documentclass[prl,twocolumn,preprintnumbers,amsmath,amssymb,hyperref,letterpaper]{revtex4}
\usepackage{graphicx}
\usepackage{latexsym}
\usepackage{amsmath}
\usepackage{graphics}
\usepackage{amssymb}
\usepackage{layout}
\usepackage{verbatim}
\usepackage{amsfonts,epsfig}
\usepackage{slashed}
\usepackage{feynmp}
\usepackage{float}
\usepackage{bm}
\usepackage{bbm}
\usepackage{hyperref}
\usepackage{epstopdf}
\newcommand{\ket}[1]{\left|#1\right>}
\newcommand{\bra}[1]{\left< #1 \right|}
\newcommand{\beq}{\begin{equation}}
\newcommand{\eeq}{\end{equation}}
\newcommand{\bea}{\begin{eqnarray}}
\newcommand{\eea}{\end{eqnarray}}
\newcommand{\nn}{\nonumber}
\newcommand{\tr}{\hbox{Tr}}

\begin{document}

\unitlength = 1mm

\title{Analytical approach to swift nonleaky entangling gates in superconducting qubits}
%cavity-coupled, microwave, leakage, non-leaky, simple, spectral crowding
\author{Sophia E. Economou$^1$ and Edwin Barnes$^{2}$}

\affiliation{$^{1}$Naval Research Laboratory, Washington, DC 20375, USA\\
$^{2}$Condensed Matter Theory Center and Joint Quantum Institute, Department of
Physics, University of Maryland, College Park, Maryland 20742-4111, USA}

\begin{abstract}
We develop schemes for designing pulses that implement fast and precise entangling quantum gates in superconducting qubit systems despite the presence of nearby harmful transitions. Our approach is based on purposely involving the nearest harmful transition in the quantum evolution instead of trying to avoid it.  Using analytical tools, we design simple microwave control fields that implement maximally entangling gates with fidelities exceeding 99\% in times as low as 40 ns. We demonstrate our approach in a two-qubit circuit QED system by designing the two most important quantum entangling gates: a conditional-NOT gate and a conditional-Z gate. Our results constitute an important step toward overcoming the problem of spectral crowding, one of the primary challenges in controlling multi-qubit systems.
\end{abstract}

\maketitle

Quantum information and quantum computing hold great promise for enabling algorithms that have no efficient classical analogs. Over the last decade, superconducting qubits have demonstrated rapid progress in coherence times, manipulation techniques, readout schemes and circuit cavity architectures \cite{Clarke_Nature08,You_Nature11,Devoret_Science13,Martinis_arXiv14}. These systems are now at the forefront of the quest for coherent many-qubit devices. Many key experiments have been carried out, including two- \cite{Chow_PRL12,Chow_NJP13,Barends_Nature14} and three-qubit \cite{DiCarlo_nature10,Fedorov_Nature12} gates, as well as proof-of-principle demonstrations of elementary quantum algorithms \cite{DiCarlo_nature09,Mariantoni_science11,Lucero_NP12} and error correction \cite{Chow_NC14,Reed_nature12}. As the number of qubits grows and the algorithms become more involved, the need for fast, high fidelity quantum control also intensifies.

There are two ways to carry out quantum gates in superconducting qubit systems. One way is to tune parameters, such as energy levels, so that the system is brought close to an anticrossing and held idle until different states accumulate the relative phases appropriate for a desired operation. This method has been used in many proposals and experimental demonstrations \cite{Strauch_PRL03,DiCarlo_nature09,DiCarlo_nature10,Mariantoni_science11,Reed_nature12,Lucero_NP12} and has recently been refined to produce faster and more precise tuning-based gates \cite{Ghosh_PRA13,Egger_arXiv13,Martinis_PRA14,Barends_Nature14}. Alternatively, external microwave pulses can be used to implement gates by exciting a prescribed set of target transitions \cite{Majer_Nature07,Li_PRB08,Kelly_PRL10,Rigetti_PRB10,Yang_PRA10,Kim_PRL11,Chow_PRL12,Fedorov_Nature12,Chow_NJP13,Chow_NC14}. Each method has merits and drawbacks, and future devices will likely employ a combination of techniques, so it is crucial to push the limits of both approaches in terms of gate speeds and fidelities.

Microwave driving allows the system to remain in the optimized parameter regime where decoherence and decay rates are low. The major obstacle to fast and high-fidelity control is the so-called spectral crowding problem, in which transitions close in frequency to the target transitions of a given gate are unintentionally excited, leading to phase and leakage errors. The obvious way to deal with this issue is to make the control pulse spectrally narrow; this translates to longer pulses in the time domain, which is not desirable as it means a greater susceptibility to decay and decoherence. To date, there have been several works that address this problem in the context of single-qubit gates by devising pulses that avoid the harmful transitions, either by brute-force numerical pulse shaping \cite{Safaei_PRB09,Rebentrost_09} or by engineering the pulse spectrum to contain sharp holes at the frequencies of the harmful transitions \cite{Motzoi_PRL09,Forney_PRA10,Gambetta_PRA11,Motzoi_PRA13,Schutjens_PRA13}. The design of microwave control protocols that combat leakage in the case of two-qubit entangling gates has remained an open problem.

\begin{figure}
\begin{center}
\includegraphics[width=0.4\columnwidth]{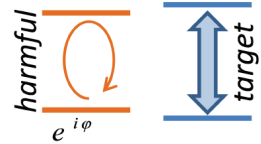}
\caption{\label{fig:cartoon} (Color online) Quantum gates can be sped up by intentionally driving nearby harmful transitions rather than avoiding them. The driving is engineered so that these transitions undergo cyclic evolution and acquire phases, which are included in the gate design.}
\end{center}
\end{figure}
In this Rapid Communication, we present an analytical approach to designing fast, high-fidelity two-qubit gates implemented with microwave pulses. The key idea is to speed up the gates by purposely driving the nearest harmful transition and incorporating its dynamics into the gate design rather than by attempting to avoid this transition. We engineer the driving so that the harmful transition subspace undergoes cyclic evolution (see Fig.~\ref{fig:cartoon}), which minimizes leakage errors and significantly enhances gate fidelities. We refer to our protocol as Speeding up Waveforms by Inducing Phases to Harmful Transitions (SWIPHT), and we design control pulses for the two most common maximally entangling gates, the conditional-$NOT$ ($CNOT$) and the conditional-$Z$ ($CZ$) gates. We employ analytical methods in order to develop a general approach that can be easily adapted to different experimental setups and other types of gates while at the same time affording transparency and insight into the dynamics of the system. We show that with our SWIPHT method, maximally entangling gates of durations 40 ns to 250 ns and fidelities ranging from 99\% to 99.94\% can be achieved with simple pulse waveforms.

We consider two superconducting transmon qubits \cite{Koch_PRA07} with finite anharmonicity coupled to the same cavity. The qubit states are the two lowest levels, $|0\rangle$ and $|1\rangle $. Additional levels outside the qubit subspace $|2\rangle, |3\rangle , ...$ are always present and, as we will show, can sometimes be useful to incorporate into gate designs. The Hamiltonian for the qubits and cavity is given by
\begin{eqnarray}
&&H_0 = \omega_c a^\dag a + \sum_{j,n} \Big[\epsilon_{j,n}\ket{j,n}\bra{j,n} \nonumber \\
&&+ g\sqrt{n{+}1}\left(a^\dag \ket{j,n}\bra{j,n{+}1}+ a \ket{j,n{+}1}\bra{j,n}\right)\Big],\label{ham}
\end{eqnarray}
where $\epsilon_{j,n}$ is the energy of the $n$th level of qubit $j$, $\omega_c$ is the cavity frequency, $g$ is the qubit-cavity coupling, and $a^\dagger$ ($a$) is the cavity mode creation (annihilation) operator. The single-qubit spectrum is given by ${\epsilon_{j,n}{=}n\epsilon_{j,1}{-}n(n{-}1)\eta/2}$, where the qubit subspace of qubit $j$ has energy splitting $\epsilon_{j,1}$, and $\eta$ is the anharmonicity. While we assume identical qubit-cavity couplings for simplicity, our results easily generalize to the case of distinct couplings for each qubit. We define the logical two-qubit states, $\ket{00}$, $\ket{01}$, $\ket{10}$, $\ket{11}$, to be states in the interacting spectrum obtained by diagonalizing $H_0$. When $g$ is the smallest energy scale in the problem and $\omega_c$ is sufficiently detuned from $\epsilon_{j,1}$, these two-qubit states, along with the lowest-energy excited states, are close to their non-interacting counterparts, so that the transitions among these states come in nearly-degenerate pairs. An external microwave field excites transitions among the interacting states through
\beq
H_p = E(t)e^{i\omega_pt}\Big[\lambda a+\sum_{j,n}\lambda_{j}\sqrt{n{+}1}\ket{j,n}\bra{j,n{+}1}\Big]+\hbox{h.c.},\nn
\eeq
where $E(t)$ denotes the magnitude of the applied electric field (with all dipole moments equal to unity) and $\omega_p$ the microwave pulse frequency. Depending on the experimental setup, it is possible to drive all qubits and the cavity or some subset of these components; we account for this by including the parameters $\lambda$, $\lambda_j$, which can take the values 0 or 1. Here, we focus on the case where only one qubit is driven, $\lambda{=}\lambda_1{=}0$, $\lambda_2{=}1$. Because of the structure of the spectrum in the small $g$ regime, a chosen target transition will typically have a unique nearest harmful transition in close spectral proximity.

The general goal of this work is to engineer pulses that generate an evolution of the form $U_{target}\oplus \mathbbm{1}$, where typically the desired evolution is nontrivial in a $2{\times}2$ subspace defined by the target transition. To achieve this evolution without resorting to spectrally selective pulses, which require resolving the frequency difference between the target and the nearest harmful transition, we allow the latter to acquire a phase $e^{2i\pi}$ so that it actively participates in the dynamics. This is a challenging problem as it requires us to solve exactly and engineer the evolutions of two time-dependent two-level systems driven simultaneously by the same pulse. We accomplish this in two ways: by adapting a recently developed reverse-engineering approach to solving the Schr\"odinger equation, and by employing analytically solvable pulse shapes with special properties. For the harmful transition we are essentially devising transitionless driving, which is in itself an interesting topic \cite{Berry_JPA09} that has found applications recently in different contexts of quantum control \cite{Li_PRA11,Fasihi_JPSJ12,Jing_PRA13,Vacanti_NJP14}.

We first focus on the most well-known entangling gate, the two-qubit $CNOT$ gate, in which the state of one qubit is either flipped or left alone depending on the state of the other qubit. We make use of a recently developed analytical method for finding pulses that exactly implement a desired evolution \cite{Barnes_PRL12,Barnes_PRA13}. We use the notation of Ref.~\cite{Barnes_PRA13}, which shows that for a two-level system driven by a pulse with detuning $\Delta$ and Rabi frequency $\Omega(t){=}dE(t)$ (where $d$ is the dipole moment of the system), both the pulse and its corresponding evolution operator $U(t)$ in the rotating frame of the driving field can be expressed in terms of a single real function, $\chi(t)$:
\bea
\Omega(t)&\!\!\!=\!\!\!&\frac{\ddot\chi}{2\sqrt{\tfrac{\Delta^2}{4}-\dot\chi^2}}-\sqrt{\tfrac{\Delta^2}{4}-\dot\chi^2}\cot(2\chi),\nn\\
U(t)&\!\!\!=\!\!\!&e^{-i\tfrac{\pi}{4}\sigma_y}\left(\begin{matrix}\cos\chi e^{i\psi_-} & \sin\chi e^{-i\psi_+} \cr -\sin\chi e^{i\psi_+} & \cos\chi e^{-i\psi_-}\end{matrix}\right),\label{chiformalism}\\
\psi_\pm(t)&\!\!\!=\!\!\!&\int_0^tdt'\sqrt{\tfrac{\Delta^2}{4}{-}\dot\chi^2(t')}\csc(2\chi(t'))\pm\tfrac{1}{2}\arcsin(\tfrac{2\dot\chi(t)}{\Delta}).\nn
\eea
The key result of Ref.~\cite{Barnes_PRA13} is that any choice of $\chi(t)$ such that the inequality $|\dot\chi|{\le}|\tfrac{\Delta}{2}|$ is satisfied yields an exact solution to the Schr\"odinger equation given by Eq.~(\ref{chiformalism}).

\begin{figure}
\begin{center}
\includegraphics[width=\columnwidth]{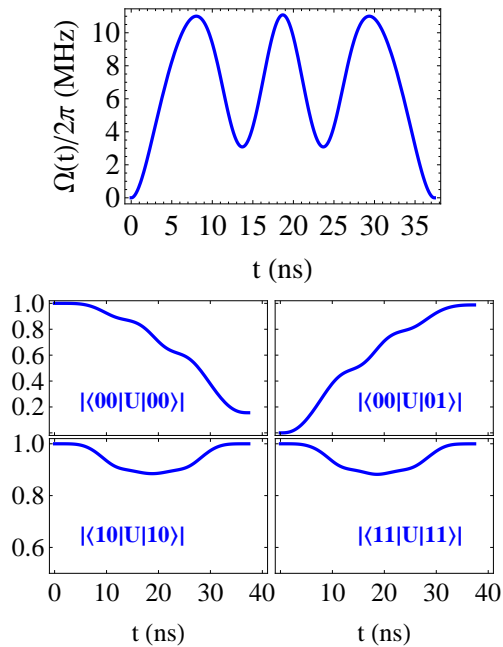}
\caption{\label{fig:cnotpulse} (Color online) Upper panel: a non-adiabatic $\pi$-pulse that induces a trivial phase on a nearby harmful transition. It implements a generalized $CNOT$ gate in 38 ns with 99\% fidelity for $\omega_c{=}7.15$ GHz, $\epsilon_{1,1}{=}6.2$ GHz, $\epsilon_{2,1}{=}6.8$ GHz, $\eta{=}350$ MHz, $g{=}250$ MHz, $\Delta{=}24.5$ MHz. Three cavity and four qubit levels are kept in numerical simulations to yield converged results. The lower four panels show the evolution in the two-qubit subspace, which is as expected for a $CNOT$ gate.}
\end{center}
\end{figure}
The formalism of Eq.~(\ref{chiformalism}) can be used for a SWIPHT-based $CNOT$ gate if we suppose that $U(t)$ in that equation refers to the evolution operator in the subspace of the harmful transition. We take the pulse to be resonant with the target transition, so that the detuning $\Delta$ is the difference of the resonance frequencies of the target and harmful transitions. The goal is then to choose $\chi(t)$ in such a way that we obtain a fast $\pi$-pulse (with duration $\tau_p$) for $\Omega(t)$ (i.e., $2\int_0^{\tau_p}dt\Omega(t){=}\pi$) while making sure that the evolution for the harmful transition at the end of the pulse coincides with the evolution that would occur in the absence of any pulse, namely we want $U(\tau_p){=}e^{i\tfrac{\Delta}{2}\tau_p\sigma_z}$. This condition can be achieved if we impose $\chi(\tau_p){=}\tfrac{\pi}{4}$, $\dot\chi(\tau_p){=}0$, and $\psi_\pm(\tau_p){=}\tfrac{\Delta\tau_p}{2}$ \cite{foot2}. Our choice of $\chi$ must also satisfy $|\dot\chi|{\le}|\tfrac{\Delta}{2}|$ and the initial conditions $\chi(0){=}\tfrac{\pi}{4}$, $\dot\chi(0){=}0$. An ansatz for $\chi$ which automatically satisfies the constraints on $\chi$ and $\dot\chi$ at $t{=}0,\tau_p$ is
\beq
\chi(t)=A(t/\tau_p)^4(1-t/\tau_p)^4+\pi/4.\label{chiansatz}
\eeq
An appropriate choice of parameters: $A{=}138.9$ and $\tau_p{=}5.87/|\Delta|$, guarantees that the remaining two conditions, $2\int_0^{\tau_p}dt\Omega(t){=}\pi$ and $\psi_\pm(\tau_p){=}\tfrac{\Delta\tau_p}{2}$, are also satisfied. Plugging the resulting $\chi$ into the formula for $\Omega(t)$ in Eq.~(\ref{chiformalism}), we obtain a $\pi$-pulse (see Fig.~\ref{fig:cnotpulse}) that implements the identity operation on the harmful transition. Since the pulse duration $\tau_p$ is inversely proportional to the frequency separation $\Delta$ of the target and harmful transitions, the gate speed is set by the cavity coupling and frequency. To verify the efficacy of this approach, we ran multi-level numerical simulations, considering the case where only the second qubit is driven, and treating $\ket{00}{\Leftrightarrow}\ket{01}$ and $\ket{10}{\Leftrightarrow}\ket{11}$ as the target and harmful transitions, respectively. For a cavity coupling of $g=250$ MHz, our simulations show that the gate time for a generalized $CNOT$ with arbitrary phases $\phi_\mu$ \cite{foot1},
\beq
CNOT=\left[\begin{matrix} 0 & e^{i\phi_a} & 0 & 0 \cr e^{i\phi_b} & 0 & 0 & 0 \cr 0 & 0 & e^{i\phi_c} & 0 \cr 0 & 0 & 0 & e^{i\phi_d} \end{matrix}\right],\label{gencnot}
\eeq
can be reduced to as low as 38 ns while maintaining a fidelity of 99\%, where we define fidelity as in Ref.~[\onlinecite{Pedersen_PLA07}]: $f\equiv\tfrac{1}{20}(\tr[UU^\dagger]+|\tr[U^\dagger CNOT]|^2)$ with $U$ the evolution operator truncated to the logical two-qubit subspace, and we optimize over $\phi_\mu$. This value is already at the threshold of surface codes \cite{Fowler_PRA12} and could be further improved by optimizing over system parameters. The fact that such high fidelities can be achieved even for this relatively strong value of the qubit-cavity coupling indicates that our approach is robust throughout the range of physical parameter values.

Another commonly used, maximally entangling gate is the $CZ$. This gate is locally equivalent to $CNOT$, but is also important in its own right, for example in the generation of cluster states for measurement-based quantum computing \cite{Raussendorf_PRA03}. It can be defined as the matrix $CZ{=}\hbox{diag}(1,-1,1,1)$, i.e., one of the two-qubit basis states (here state $|01\rangle$) acquires a phase $\pi$ while the rest remain unaffected by the control. It is natural to construct such gates using cyclic, phase-inducing evolutions that involve states outside the two-qubit subspace. To implement our SWIPHT protocol we again require that the target transition, $\ket{01}\Leftrightarrow\ket{02}$, acquires a phase of $\pi$ while the nearest harmful transition, $\ket{11}\Leftrightarrow\ket{12}$, acquires a phase of $2\pi$. To achieve the latter, we recall one of the properties of SU(2): undergoing two cycles of cyclic evolution resonantly is equivalent to the identity operation. Thus, any resonant waveform with pulse area $4\pi$ acting on the harmful transition will be equivalent to the identity operation on the basis states involved in the transition. The remaining challenge is of course to select the pulse shape and remaining pulse parameters such that the desired evolution (the minus sign) is induced on the states comprising the target transition.

\begin{figure*}
\begin{center}
\includegraphics[width=2\columnwidth]{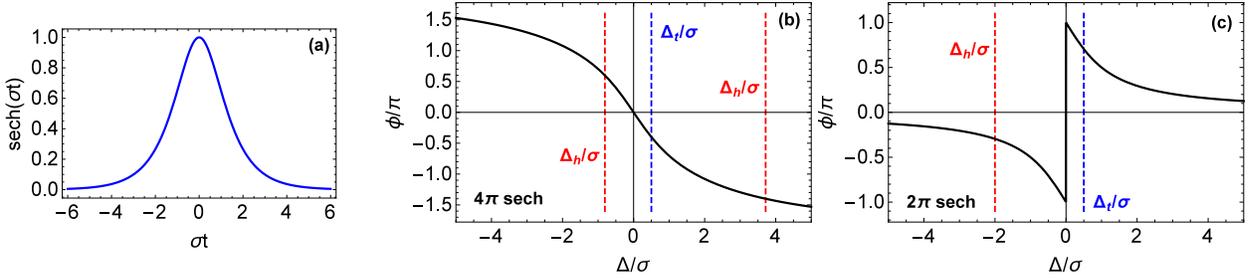}
\caption{\label{fig:sechphases} (Color online) (a) Hyperbolic secant pulse. (b) Phase induced by the $4\pi$ and (c) $2\pi$ hyperbolic secant pulses versus the ratio of detuning $\Delta$ to pulse bandwidth $\sigma$. A generalized $CZ$ gate can be achieved by inducing phases on both the target and nearest harmful transitions such that their difference is $\pm\pi$. For a given choice of the target's ratio $\Delta_t/\sigma$ (blue vertical line), there are two solutions for the harmful transition's ratio $\Delta_h/\sigma$ (red vertical lines) in the case of the $4\pi$ sech, but only one solution in the case of the $2\pi$ sech.}
\end{center}
\end{figure*}
We solve this problem by using a hyperbolic secant (sech) envelope, $\Omega(t){=}\Omega_0 \text{sech}(\sigma t)$, shown in Fig.~\ref{fig:sechphases}a. This pulse yields analytically solvable two-level system dynamics \cite{Rosen_PR32}, and moreover the fully transitionless condition is fulfilled by choosing $\Omega_0/\sigma$ appropriately, \emph{independently} of the detuning \cite{Economou_PRB06}. In addition, the sech pulse is a particularly simple waveform that is easily generated in the laboratory. The phase induced on the states of a transition by a 4$\pi$ sech pulse is given by \cite{Economou_PRB12}
\begin{eqnarray}
\phi=2\arctan\left[\frac{4\Delta/\sigma}{\left(\Delta/\sigma\right)^2-3} \right],
\label{phase4pisech}
\end{eqnarray}
where $\Delta$ is the detuning, i.e., $\Delta{=}\omega_p{-}\omega$ for a transition of frequency $\omega$. This phase is shown in Fig.~\ref{fig:sechphases}b as a function of the ratio $\Delta/\sigma$. By making the pulse resonant with the harmful transition ($\omega_p{=}\omega_h$) and choosing $\Omega_0/\sigma=2$, we guarantee that this transition acquires a trivial $e^{2i\pi}$ phase. We still have freedom to choose $\sigma$ so that a $\pi$ phase can be induced on the target transition by setting $\sigma{=}|\Delta_t|/\sqrt{3}$, where $\Delta_t{=}\omega_t{-}\omega_h{\equiv}\delta\omega$ is now the frequency difference of the two transitions.

\begin{figure}
\begin{center}
\includegraphics[width=0.8\columnwidth]{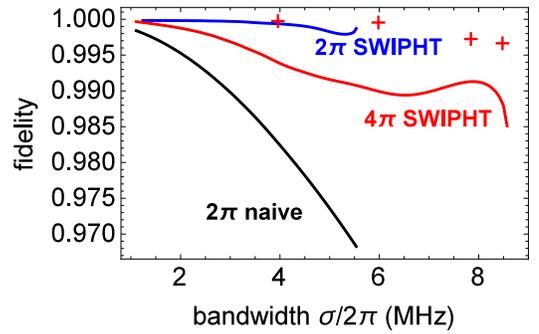}
\caption{\label{fig:CZfidelities} (Color online) Fidelity versus bandwidth for the parameters of Fig.~\ref{fig:cnotpulse} but with $g{=}130$ MHz, $\omega_t{=}6.447$ GHz, $\omega_h{=}6.458$ GHz for the $2\pi$ SWIPHT (Eq.~(\ref{wpfroms2pi})), $4\pi$ SWIPHT (Eq.~(\ref{wpfroms4pi})), and $2\pi$ naive (with $\omega_p{=}\omega_t$) pulses. Solid curves are not optimized over system or pulse parameters. Crosses indicate $4\pi$ SWIPHT gates optimized over pulse parameters.}
\end{center}
\end{figure}

We in fact have significantly more flexibility in designing the gate, as revealed in Fig.~\ref{fig:sechphases}b, where it is apparent that there are infinitely many ways in which the \emph{difference} in the phases acquired by the target and harmful transitions is equal to $\pm\pi$. Allowing both transitions to acquire a phase generates a generalized $CZ$ gate, defined as
$\widetilde{CZ}{=}\hbox{diag}(e^{i\phi_{00}},e^{i\phi_{01}},e^{i\phi_{10}},e^{i\phi_{11}})$ with $\phi_{00}{-}\phi_{01}{-}\phi_{10}{+}\phi_{11}{=}{\pm}\pi$. This gate is completely equivalent to $CZ$ \cite{foot3}. As shown in Fig.~\ref{fig:sechphases}b, for a given choice of the detuning from the target, there are in general two solutions for $\Delta_h/\sigma$ which produce the correct phase difference. Since $\Delta_t$ and $\Delta_h$ are related by $\Delta_t{=}\Delta_h{-}\delta\omega$, where the difference in the transition frequencies $\delta\omega$ is fixed by the system parameters $g,\eta,\omega_c$ and independent of the pulse, each solution can be interpreted as fixing the pulse frequency in terms of the bandwidth \cite{foot2}:
\beq
\omega_p=\frac{\omega_t{+}\omega_h}{2}{\pm}\frac{1}{2}\sqrt{\delta\omega^2{-}20\sigma^2{\pm}4\sigma\sqrt{16 \sigma ^2{+}3 \delta\omega^2}}.\label{wpfroms4pi}
\eeq
All together there are four distinct solutions for each value of $\sigma$, one for each choice of the signs inside and outside the square root; in addition to the two choices of $\Delta_h/\sigma$ evident in Fig.~\ref{fig:sechphases}b, two more solutions arise from interchanging how the target and harmful transitions are defined. Eq.~(\ref{wpfroms4pi}) reveals that there is a maximal value of $\sigma$ beyond which $\omega_p$ becomes unphysical: $\sigma_{max}{=}\tfrac{2{+}\sqrt{7}}{6}|\delta\omega|$, which corresponds to setting the pulse frequency halfway between the two transitions: $\omega_p{=}(\omega_t{+}\omega_h)/2$. Comparing $\sigma_{max}$ with the value of $\sigma$ for an ordinary $CZ$, $\sigma{=}|\delta\omega|/\sqrt{3}$, we see that the gate speed of a generalized $CZ$ can be up to 34\% faster. Fig.~\ref{fig:CZfidelities} shows the fidelity of the $\widetilde{CZ}$ gate as a function of pulse bandwidth. For the parameters used in the figure, the fastest pulse is 185 ns. Without any optimization with respect to either system or pulse parameters, the fidelity ranges from 98.5\% to 99.95\%. The deviation from perfect fidelities comes from the presence of additional harmful transitions as well as from the fact that the target and nearest harmful transitions have different dipole moments. Further optimization over the pulse parameters $\Omega_0$ and $\omega_p$ reduces these effects, yielding significantly higher fidelities ranging from 99.7\% to 99.99\% (see Fig.~\ref{fig:CZfidelities}). Additional optimization over system parameters should produce even better fidelities.

For the generalized gate, we can improve the fidelity still further by using a much weaker pulse. A 2$\pi$ sech pulse induces a phase $\phi{=}2\arctan\left(\sigma/\Delta\right)$ \cite{Economou_PRB12}, shown in Fig.~\ref{fig:sechphases}c. As before, we may choose a value of $\Delta_t/\sigma$ for the target transition, and the corresponding value for the harmful transition is determined by requiring a phase difference of $\pm\pi$. However, unlike in the $4\pi$ sech case, here there is only one solution for $\Delta_h/\sigma$ once $\Delta_t/\sigma$ is fixed (see Fig.~\ref{fig:sechphases}c). We then obtain two solutions for the pulse frequency for each value of $\sigma$:
\beq
\omega_p=\frac{\omega_t{+}\omega_h}{2}\pm\frac{1}{2}\sqrt{\delta\omega^2-4\sigma^2}.\label{wpfroms2pi}
\eeq
The maximal bandwidth is again attained for $\omega_p{=}(\omega_t{+}\omega_h)/2$, but now at a value of $\sigma_{max}{=}|\delta\omega|/2$, showing that there is a cost in terms of gate speed when the weaker $2\pi$ sech pulse is used. Fig.~\ref{fig:CZfidelities} shows fidelity as a function of bandwidth for the SWIPHT $\widetilde{CZ}$ gate for both the $2\pi$ and $4\pi$ pulses, along with a ``naive" gate which is implemented using a $2\pi$ sech pulse which is held resonant with the target transition for all values of $\sigma$ and so does not purposely involve the nearest harmful transition in the gate design. The figure shows a striking improvement in gate fidelities when the SWIPHT protocol is used even without optimizing over system or pulse parameters, particularly at higher gate speeds (large $\sigma$). The $4\pi$ SWIPHT pulse extends to larger values of $\sigma$, so that although one gains in fidelity by using the $2\pi$ SWIPHT pulse, faster gate speeds can be achieved with the $4\pi$ SWIPHT pulse.

In conclusion, we have developed a versatile analytical protocol for creating fast high fidelity gates by incorporating into the quantum dynamics the nearest harmful transition, which is typically avoided in prior quantum control schemes. We have demonstrated the effectiveness of our approach by designing simple, smooth pulses that implement the two most common maximally entangling gates, the $CNOT$ and $CZ$, showing that the speeds of microwave-driven gates can be substantially increased while keeping fidelities above the threshold of error correcting codes. We expect that our methods will serve as a powerful tool in achieving high fidelity control of large scale, multiqubit systems.

This work was supported by LPS-CMTC (EB) and in part by ONR (SEE).


\begin{thebibliography}{45}
\expandafter\ifx\csname natexlab\endcsname\relax\def\natexlab#1{#1}\fi
\expandafter\ifx\csname bibnamefont\endcsname\relax
  \def\bibnamefont#1{#1}\fi
\expandafter\ifx\csname bibfnamefont\endcsname\relax
  \def\bibfnamefont#1{#1}\fi
\expandafter\ifx\csname citenamefont\endcsname\relax
  \def\citenamefont#1{#1}\fi
\expandafter\ifx\csname url\endcsname\relax
  \def\url#1{\texttt{#1}}\fi
\expandafter\ifx\csname urlprefix\endcsname\relax\def\urlprefix{URL }\fi
\providecommand{\bibinfo}[2]{#2}
\providecommand{\eprint}[2][]{\url{#2}}

\bibitem[{\citenamefont{Clarke and Wilhelm}(2008)}]{Clarke_Nature08}
\bibinfo{author}{\bibfnamefont{J.}~\bibnamefont{Clarke}} \bibnamefont{and}
  \bibinfo{author}{\bibfnamefont{F.~K.} \bibnamefont{Wilhelm}},
  \bibinfo{journal}{Nature} \textbf{\bibinfo{volume}{453}},
  \bibinfo{pages}{1031} (\bibinfo{year}{2008}).

\bibitem[{\citenamefont{You and Nori}(2011)}]{You_Nature11}
\bibinfo{author}{\bibfnamefont{J.~Q.} \bibnamefont{You}} \bibnamefont{and}
  \bibinfo{author}{\bibfnamefont{F.}~\bibnamefont{Nori}},
  \bibinfo{journal}{Nature} \textbf{\bibinfo{volume}{474}},
  \bibinfo{pages}{589} (\bibinfo{year}{2011}).

\bibitem[{\citenamefont{Devoret and Schoelkopf}(2013)}]{Devoret_Science13}
\bibinfo{author}{\bibfnamefont{M.~H.} \bibnamefont{Devoret}} \bibnamefont{and}
  \bibinfo{author}{\bibfnamefont{R.~J.} \bibnamefont{Schoelkopf}},
  \bibinfo{journal}{Science} \textbf{\bibinfo{volume}{339}},
  \bibinfo{pages}{1169} (\bibinfo{year}{2013}).

\bibitem[{\citenamefont{Martinis and Megrant}(2014)}]{Martinis_arXiv14}
\bibinfo{author}{\bibfnamefont{J.~M.} \bibnamefont{Martinis}} \bibnamefont{and}
  \bibinfo{author}{\bibfnamefont{A.}~\bibnamefont{Megrant}},
  \bibinfo{journal}{arXiv:1410.5793}  (\bibinfo{year}{2014}).

\bibitem[{\citenamefont{Chow et~al.}(2012)\citenamefont{Chow, Gambetta,
  Corcoles, Merkel, Smolin, Rigetti, Poletto, Keefe, Rothwell, Rozen
  et~al.}}]{Chow_PRL12}
\bibinfo{author}{\bibfnamefont{J.~M.} \bibnamefont{Chow}},
  \bibinfo{author}{\bibfnamefont{J.~M.} \bibnamefont{Gambetta}},
  \bibinfo{author}{\bibfnamefont{A.~D.} \bibnamefont{Corcoles}},
  \bibinfo{author}{\bibfnamefont{S.~T.} \bibnamefont{Merkel}},
  \bibinfo{author}{\bibfnamefont{J.~A.} \bibnamefont{Smolin}},
  \bibinfo{author}{\bibfnamefont{C.}~\bibnamefont{Rigetti}},
  \bibinfo{author}{\bibfnamefont{S.}~\bibnamefont{Poletto}},
  \bibinfo{author}{\bibfnamefont{G.~A.} \bibnamefont{Keefe}},
  \bibinfo{author}{\bibfnamefont{M.~B.} \bibnamefont{Rothwell}},
  \bibinfo{author}{\bibfnamefont{J.~R.} \bibnamefont{Rozen}},
  \bibnamefont{et~al.}, \bibinfo{journal}{Phys.\ Rev.\ Lett.}
  \textbf{\bibinfo{volume}{109}}, \bibinfo{pages}{060501}
  (\bibinfo{year}{2012}).

\bibitem[{\citenamefont{Chow et~al.}(2013)\citenamefont{Chow, Gambetta, Cross,
  Merkel, Rigetti, and Steffen}}]{Chow_NJP13}
\bibinfo{author}{\bibfnamefont{J.~M.} \bibnamefont{Chow}},
  \bibinfo{author}{\bibfnamefont{J.~M.} \bibnamefont{Gambetta}},
  \bibinfo{author}{\bibfnamefont{A.~W.} \bibnamefont{Cross}},
  \bibinfo{author}{\bibfnamefont{S.~T.} \bibnamefont{Merkel}},
  \bibinfo{author}{\bibfnamefont{C.}~\bibnamefont{Rigetti}}, \bibnamefont{and}
  \bibinfo{author}{\bibfnamefont{M.}~\bibnamefont{Steffen}},
  \bibinfo{journal}{New\ J.\ Phys.} \textbf{\bibinfo{volume}{15}},
  \bibinfo{pages}{115012} (\bibinfo{year}{2013}).

\bibitem[{\citenamefont{Barends et~al.}(2014)\citenamefont{Barends, Kelly,
  Megrant, Veitia, Sank, Jeffrey, White, Mutus, Fowler, Campbell
  et~al.}}]{Barends_Nature14}
\bibinfo{author}{\bibfnamefont{R.}~\bibnamefont{Barends}},
  \bibinfo{author}{\bibfnamefont{J.}~\bibnamefont{Kelly}},
  \bibinfo{author}{\bibfnamefont{A.}~\bibnamefont{Megrant}},
  \bibinfo{author}{\bibfnamefont{A.}~\bibnamefont{Veitia}},
  \bibinfo{author}{\bibfnamefont{D.}~\bibnamefont{Sank}},
  \bibinfo{author}{\bibfnamefont{E.}~\bibnamefont{Jeffrey}},
  \bibinfo{author}{\bibfnamefont{T.~C.} \bibnamefont{White}},
  \bibinfo{author}{\bibfnamefont{J.}~\bibnamefont{Mutus}},
  \bibinfo{author}{\bibfnamefont{A.~G.} \bibnamefont{Fowler}},
  \bibinfo{author}{\bibfnamefont{B.}~\bibnamefont{Campbell}},
  \bibnamefont{et~al.}, \bibinfo{journal}{Nature}
  \textbf{\bibinfo{volume}{508}}, \bibinfo{pages}{500} (\bibinfo{year}{2014}).

\bibitem[{\citenamefont{DiCarlo et~al.}(2010)\citenamefont{DiCarlo, Reed, Sun,
  Johnson, Chow, Gambetta, Frunzio, Girvin, Devoret, and
  Schoelkopf}}]{DiCarlo_nature10}
\bibinfo{author}{\bibfnamefont{L.}~\bibnamefont{DiCarlo}},
  \bibinfo{author}{\bibfnamefont{M.~D.} \bibnamefont{Reed}},
  \bibinfo{author}{\bibfnamefont{L.}~\bibnamefont{Sun}},
  \bibinfo{author}{\bibfnamefont{B.~R.} \bibnamefont{Johnson}},
  \bibinfo{author}{\bibfnamefont{J.}~\bibnamefont{Chow}},
  \bibinfo{author}{\bibfnamefont{J.~M.} \bibnamefont{Gambetta}},
  \bibinfo{author}{\bibfnamefont{L.}~\bibnamefont{Frunzio}},
  \bibinfo{author}{\bibfnamefont{S.~M.} \bibnamefont{Girvin}},
  \bibinfo{author}{\bibfnamefont{M.~H.} \bibnamefont{Devoret}},
  \bibnamefont{and} \bibinfo{author}{\bibfnamefont{R.~J.}
  \bibnamefont{Schoelkopf}}, \bibinfo{journal}{Nature}
  \textbf{\bibinfo{volume}{467}}, \bibinfo{pages}{574} (\bibinfo{year}{2010}).

\bibitem[{\citenamefont{Fedorov et~al.}(2012)\citenamefont{Fedorov, Steffen,
  Baur, da~Silva, and A.Wallraff}}]{Fedorov_Nature12}
\bibinfo{author}{\bibfnamefont{A.}~\bibnamefont{Fedorov}},
  \bibinfo{author}{\bibfnamefont{L.}~\bibnamefont{Steffen}},
  \bibinfo{author}{\bibfnamefont{M.}~\bibnamefont{Baur}},
  \bibinfo{author}{\bibfnamefont{M.~P.} \bibnamefont{da~Silva}},
  \bibnamefont{and} \bibinfo{author}{\bibnamefont{A.Wallraff}},
  \bibinfo{journal}{Nature} \textbf{\bibinfo{volume}{481}},
  \bibinfo{pages}{170} (\bibinfo{year}{2012}).

\bibitem[{\citenamefont{DiCarlo et~al.}(2009)\citenamefont{DiCarlo, Chow,
  Gambetta, Bishop, B.~R.~Johnson, Majer, Blais, Frunzio, Girvin, and
  Schoelkopf}}]{DiCarlo_nature09}
\bibinfo{author}{\bibfnamefont{L.}~\bibnamefont{DiCarlo}},
  \bibinfo{author}{\bibfnamefont{J.~M.} \bibnamefont{Chow}},
  \bibinfo{author}{\bibfnamefont{J.~M.} \bibnamefont{Gambetta}},
  \bibinfo{author}{\bibfnamefont{L.~S.} \bibnamefont{Bishop}},
  \bibinfo{author}{\bibfnamefont{D.~I.~S.} \bibnamefont{B.~R.~Johnson}},
  \bibinfo{author}{\bibfnamefont{J.}~\bibnamefont{Majer}},
  \bibinfo{author}{\bibfnamefont{A.}~\bibnamefont{Blais}},
  \bibinfo{author}{\bibfnamefont{L.}~\bibnamefont{Frunzio}},
  \bibinfo{author}{\bibfnamefont{S.~M.} \bibnamefont{Girvin}},
  \bibnamefont{and} \bibinfo{author}{\bibfnamefont{R.~J.}
  \bibnamefont{Schoelkopf}}, \bibinfo{journal}{Nature}
  \textbf{\bibinfo{volume}{460}}, \bibinfo{pages}{240} (\bibinfo{year}{2009}).

\bibitem[{\citenamefont{Mariantoni et~al.}(2011)\citenamefont{Mariantoni, Wang,
  Yamamoto, Neeley, Bialczak, Chen, Lenander, Lucero, O’Connell, Sank
  et~al.}}]{Mariantoni_science11}
\bibinfo{author}{\bibfnamefont{M.}~\bibnamefont{Mariantoni}},
  \bibinfo{author}{\bibfnamefont{H.}~\bibnamefont{Wang}},
  \bibinfo{author}{\bibfnamefont{T.}~\bibnamefont{Yamamoto}},
  \bibinfo{author}{\bibfnamefont{M.}~\bibnamefont{Neeley}},
  \bibinfo{author}{\bibfnamefont{R.~C.} \bibnamefont{Bialczak}},
  \bibinfo{author}{\bibfnamefont{Y.}~\bibnamefont{Chen}},
  \bibinfo{author}{\bibfnamefont{M.}~\bibnamefont{Lenander}},
  \bibinfo{author}{\bibfnamefont{E.}~\bibnamefont{Lucero}},
  \bibinfo{author}{\bibfnamefont{A.~D.} \bibnamefont{O’Connell}},
  \bibinfo{author}{\bibfnamefont{D.}~\bibnamefont{Sank}}, \bibnamefont{et~al.},
  \bibinfo{journal}{Science} \textbf{\bibinfo{volume}{334}},
  \bibinfo{pages}{61} (\bibinfo{year}{2011}).

\bibitem[{\citenamefont{Lucero et~al.}(2012)\citenamefont{Lucero, Barends,
  Chen, Kelly, Mariantoni, Megrant, O'Malley, Sank, Vainsencher, Wenner
  et~al.}}]{Lucero_NP12}
\bibinfo{author}{\bibfnamefont{E.}~\bibnamefont{Lucero}},
  \bibinfo{author}{\bibfnamefont{R.}~\bibnamefont{Barends}},
  \bibinfo{author}{\bibfnamefont{Y.}~\bibnamefont{Chen}},
  \bibinfo{author}{\bibfnamefont{J.}~\bibnamefont{Kelly}},
  \bibinfo{author}{\bibfnamefont{M.}~\bibnamefont{Mariantoni}},
  \bibinfo{author}{\bibfnamefont{A.}~\bibnamefont{Megrant}},
  \bibinfo{author}{\bibfnamefont{P.}~\bibnamefont{O'Malley}},
  \bibinfo{author}{\bibfnamefont{D.}~\bibnamefont{Sank}},
  \bibinfo{author}{\bibfnamefont{A.}~\bibnamefont{Vainsencher}},
  \bibinfo{author}{\bibfnamefont{J.}~\bibnamefont{Wenner}},
  \bibnamefont{et~al.}, \bibinfo{journal}{Nat.\ Phys.}
  \textbf{\bibinfo{volume}{8}}, \bibinfo{pages}{719} (\bibinfo{year}{2012}).

\bibitem[{\citenamefont{Chow et~al.}(2014)\citenamefont{Chow, Gambetta, Magesan, Abraham, Cross, Johnson, Masluk, Ryan, Smolin, Srinivasan, and Steffen}}]{Chow_NC14}
\bibinfo{author}{\bibfnamefont{J.~M.} \bibnamefont{Chow}},
  \bibinfo{author}{\bibfnamefont{J.~M.} \bibnamefont{Gambetta}},
  \bibinfo{author}{\bibfnamefont{E.}~\bibnamefont{Magesan}},
  \bibinfo{author}{\bibfnamefont{D.~W.} \bibnamefont{Abraham}},
  \bibinfo{author}{\bibfnamefont{A.~W.} \bibnamefont{Cross}},
  \bibinfo{author}{\bibfnamefont{B.~R.} \bibnamefont{Johnson}},
  \bibinfo{author}{\bibfnamefont{N.~A.} \bibnamefont{Masluk}},
  \bibinfo{author}{\bibfnamefont{C.~A.} \bibnamefont{Ryan}},
  \bibinfo{author}{\bibfnamefont{J.~A.} \bibnamefont{Smolin}},
  \bibinfo{author}{\bibfnamefont{S.~J.} \bibnamefont{Srinivasan}}, \bibnamefont{and}
  \bibinfo{author}{\bibfnamefont{M.}~\bibnamefont{Steffen}},
  \bibinfo{journal}{Nat.\ Commun.} \textbf{\bibinfo{volume}{5}},
  \bibinfo{pages}{4015} (\bibinfo{year}{2014}).

\bibitem[{\citenamefont{Reed et~al.}(2012)\citenamefont{Reed, DiCarlo, Nigg,
  Sun, Frunzio, Girvin, and Schoelkopf}}]{Reed_nature12}
\bibinfo{author}{\bibfnamefont{M.~D.} \bibnamefont{Reed}},
  \bibinfo{author}{\bibfnamefont{L.}~\bibnamefont{DiCarlo}},
  \bibinfo{author}{\bibfnamefont{S.~E.} \bibnamefont{Nigg}},
  \bibinfo{author}{\bibfnamefont{L.}~\bibnamefont{Sun}},
  \bibinfo{author}{\bibfnamefont{L.}~\bibnamefont{Frunzio}},
  \bibinfo{author}{\bibfnamefont{S.~M.} \bibnamefont{Girvin}},
  \bibnamefont{and} \bibinfo{author}{\bibfnamefont{R.~J.}
  \bibnamefont{Schoelkopf}}, \bibinfo{journal}{Nature}
  \textbf{\bibinfo{volume}{482}}, \bibinfo{pages}{382} (\bibinfo{year}{2012}).

\bibitem[{\citenamefont{Strauch et~al.}(2003)\citenamefont{Strauch, Johnson,
  Dragt, Lobb, Anderson, and Wellstood}}]{Strauch_PRL03}
\bibinfo{author}{\bibfnamefont{F.~W.} \bibnamefont{Strauch}},
  \bibinfo{author}{\bibfnamefont{P.~R.} \bibnamefont{Johnson}},
  \bibinfo{author}{\bibfnamefont{A.~J.} \bibnamefont{Dragt}},
  \bibinfo{author}{\bibfnamefont{C.}~\bibnamefont{Lobb}},
  \bibinfo{author}{\bibfnamefont{J.}~\bibnamefont{Anderson}}, \bibnamefont{and}
  \bibinfo{author}{\bibfnamefont{F.}~\bibnamefont{Wellstood}},
  \bibinfo{journal}{Phys.\ Rev.\ Lett.} \textbf{\bibinfo{volume}{91}},
  \bibinfo{pages}{167005} (\bibinfo{year}{2003}).

\bibitem[{\citenamefont{Ghosh et~al.}(2013)\citenamefont{Ghosh, Galiautdinov,
  Zhou, Korotkov, Martinis, and Geller}}]{Ghosh_PRA13}
\bibinfo{author}{\bibfnamefont{J.}~\bibnamefont{Ghosh}},
  \bibinfo{author}{\bibfnamefont{A.}~\bibnamefont{Galiautdinov}},
  \bibinfo{author}{\bibfnamefont{Z.}~\bibnamefont{Zhou}},
  \bibinfo{author}{\bibfnamefont{A.~N.} \bibnamefont{Korotkov}},
  \bibinfo{author}{\bibfnamefont{J.~M.} \bibnamefont{Martinis}},
  \bibnamefont{and} \bibinfo{author}{\bibfnamefont{M.~R.}
  \bibnamefont{Geller}}, \bibinfo{journal}{Phys.\ Rev.\ A}
  \textbf{\bibinfo{volume}{87}}, \bibinfo{pages}{022309}
  (\bibinfo{year}{2013}).

\bibitem[{\citenamefont{Egger and Wilhelm}(2013)}]{Egger_arXiv13}
\bibinfo{author}{\bibfnamefont{D.~J.} \bibnamefont{Egger}} \bibnamefont{and}
  \bibinfo{author}{\bibfnamefont{F.~K.} \bibnamefont{Wilhelm}},
  \bibinfo{journal}{arXiv:1306.6894}  (\bibinfo{year}{2013}).

\bibitem[{\citenamefont{Martinis and Geller}(2014)}]{Martinis_PRA14}
\bibinfo{author}{\bibfnamefont{J.~M.} \bibnamefont{Martinis}} \bibnamefont{and}
  \bibinfo{author}{\bibfnamefont{M.~R.} \bibnamefont{Geller}},
  \bibinfo{journal}{Phys.\ Rev.\ A} \textbf{\bibinfo{volume}{90}},
  \bibinfo{pages}{022307} (\bibinfo{year}{2014}).

\bibitem[{\citenamefont{Majer et~al.}(2007)\citenamefont{Majer, Chow, Gambetta,
  Koch, Johnson, Schreier, Frunzio, Schuster, Houck, Wallraff
  et~al.}}]{Majer_Nature07}
\bibinfo{author}{\bibfnamefont{J.}~\bibnamefont{Majer}},
  \bibinfo{author}{\bibfnamefont{J.~M.} \bibnamefont{Chow}},
  \bibinfo{author}{\bibfnamefont{J.~M.} \bibnamefont{Gambetta}},
  \bibinfo{author}{\bibfnamefont{J.}~\bibnamefont{Koch}},
  \bibinfo{author}{\bibfnamefont{B.~R.} \bibnamefont{Johnson}},
  \bibinfo{author}{\bibfnamefont{J.~A.} \bibnamefont{Schreier}},
  \bibinfo{author}{\bibfnamefont{L.}~\bibnamefont{Frunzio}},
  \bibinfo{author}{\bibfnamefont{D.~I.} \bibnamefont{Schuster}},
  \bibinfo{author}{\bibfnamefont{A.~A.} \bibnamefont{Houck}},
  \bibinfo{author}{\bibfnamefont{A.}~\bibnamefont{Wallraff}},
  \bibnamefont{et~al.}, \bibinfo{journal}{Nature}
  \textbf{\bibinfo{volume}{449}}, \bibinfo{pages}{443} (\bibinfo{year}{2007}).

\bibitem[{\citenamefont{Li et~al.}(2008)\citenamefont{Li, Chalapat, and
  Paraoanu}}]{Li_PRB08}
\bibinfo{author}{\bibfnamefont{J.}~\bibnamefont{Li}},
  \bibinfo{author}{\bibfnamefont{K.}~\bibnamefont{Chalapat}}, \bibnamefont{and}
  \bibinfo{author}{\bibfnamefont{G.~S.} \bibnamefont{Paraoanu}},
  \bibinfo{journal}{Phys.\ Rev.\ B} \textbf{\bibinfo{volume}{78}},
  \bibinfo{pages}{064503} (\bibinfo{year}{2008}).

\bibitem[{\citenamefont{Kelly et~al.}(2010)\citenamefont{Kelly, Dutton,
  Schlafer, Mookerji, Ohki, Kline, and Pappas}}]{Kelly_PRL10}
\bibinfo{author}{\bibfnamefont{W.~R.} \bibnamefont{Kelly}},
  \bibinfo{author}{\bibfnamefont{Z.}~\bibnamefont{Dutton}},
  \bibinfo{author}{\bibfnamefont{J.}~\bibnamefont{Schlafer}},
  \bibinfo{author}{\bibfnamefont{B.}~\bibnamefont{Mookerji}},
  \bibinfo{author}{\bibfnamefont{T.~A.} \bibnamefont{Ohki}},
  \bibinfo{author}{\bibfnamefont{J.~S.} \bibnamefont{Kline}}, \bibnamefont{and}
  \bibinfo{author}{\bibfnamefont{D.~P.} \bibnamefont{Pappas}},
  \bibinfo{journal}{Phys.\ Rev.\ Lett.} \textbf{\bibinfo{volume}{104}},
  \bibinfo{pages}{163601} (\bibinfo{year}{2010}).

\bibitem[{\citenamefont{Rigetti and Devoret}(2010)}]{Rigetti_PRB10}
\bibinfo{author}{\bibfnamefont{C.}~\bibnamefont{Rigetti}} \bibnamefont{and}
  \bibinfo{author}{\bibfnamefont{M.}~\bibnamefont{Devoret}},
  \bibinfo{journal}{Phys.\ Rev.\ B} \textbf{\bibinfo{volume}{81}},
  \bibinfo{pages}{134507} (\bibinfo{year}{2010}).

\bibitem[{\citenamefont{Yang et~al.}(2010)\citenamefont{Yang, Zheng, and
  Nori}}]{Yang_PRA10}
\bibinfo{author}{\bibfnamefont{C.-P.} \bibnamefont{Yang}},
  \bibinfo{author}{\bibfnamefont{S.-B.} \bibnamefont{Zheng}}, \bibnamefont{and}
  \bibinfo{author}{\bibfnamefont{F.}~\bibnamefont{Nori}},
  \bibinfo{journal}{Phys.\ Rev.\ A} \textbf{\bibinfo{volume}{82}},
  \bibinfo{pages}{062326} (\bibinfo{year}{2010}).

\bibitem[{\citenamefont{Kim et~al.}(2011)\citenamefont{Kim, Suri, Zaretskey,
  Novikov, Osborn, Mizel, Wellstood, and Palmer}}]{Kim_PRL11}
\bibinfo{author}{\bibfnamefont{Z.}~\bibnamefont{Kim}},
  \bibinfo{author}{\bibfnamefont{B.}~\bibnamefont{Suri}},
  \bibinfo{author}{\bibfnamefont{V.}~\bibnamefont{Zaretskey}},
  \bibinfo{author}{\bibfnamefont{S.}~\bibnamefont{Novikov}},
  \bibinfo{author}{\bibfnamefont{K.~D.} \bibnamefont{Osborn}},
  \bibinfo{author}{\bibfnamefont{A.}~\bibnamefont{Mizel}},
  \bibinfo{author}{\bibfnamefont{F.~C.} \bibnamefont{Wellstood}},
  \bibnamefont{and} \bibinfo{author}{\bibfnamefont{B.~S.}
  \bibnamefont{Palmer}}, \bibinfo{journal}{Phys.\ Rev.\ Lett.}
  \textbf{\bibinfo{volume}{106}}, \bibinfo{pages}{120501}
  (\bibinfo{year}{2011}).

\bibitem[{\citenamefont{Safaei et~al.}(2009)\citenamefont{Safaei, Montangero,
  Taddei, and Fazio}}]{Safaei_PRB09}
\bibinfo{author}{\bibfnamefont{S.}~\bibnamefont{Safaei}},
  \bibinfo{author}{\bibfnamefont{S.}~\bibnamefont{Montangero}},
  \bibinfo{author}{\bibfnamefont{F.}~\bibnamefont{Taddei}}, \bibnamefont{and}
  \bibinfo{author}{\bibfnamefont{R.}~\bibnamefont{Fazio}},
  \bibinfo{journal}{Phys.\ Rev.\ B} \textbf{\bibinfo{volume}{79}},
  \bibinfo{pages}{064524} (\bibinfo{year}{2009}).

\bibitem[{\citenamefont{Rebentrost and Wilhelm}(2009)}]{Rebentrost_09}
\bibinfo{author}{\bibfnamefont{P.}~\bibnamefont{Rebentrost}} \bibnamefont{and}
  \bibinfo{author}{\bibfnamefont{F.~K.} \bibnamefont{Wilhelm}},
  \bibinfo{journal}{PRB} \textbf{\bibinfo{volume}{79}},
  \bibinfo{pages}{060507(R)} (\bibinfo{year}{2009}).

\bibitem[{\citenamefont{Motzoi et~al.}(2009)\citenamefont{Motzoi, Gambetta,
  Rebentrost, and Wilhelm}}]{Motzoi_PRL09}
\bibinfo{author}{\bibfnamefont{F.}~\bibnamefont{Motzoi}},
  \bibinfo{author}{\bibfnamefont{J.~M.} \bibnamefont{Gambetta}},
  \bibinfo{author}{\bibfnamefont{P.}~\bibnamefont{Rebentrost}},
  \bibnamefont{and} \bibinfo{author}{\bibfnamefont{F.~K.}
  \bibnamefont{Wilhelm}}, \bibinfo{journal}{Phys.\ Rev.\ Lett.}
  \textbf{\bibinfo{volume}{103}}, \bibinfo{pages}{110501}
  (\bibinfo{year}{2009}).

\bibitem[{\citenamefont{Forney et~al.}(2010)\citenamefont{Forney, Jackson, and
  Strauch}}]{Forney_PRA10}
\bibinfo{author}{\bibfnamefont{A.~M.} \bibnamefont{Forney}},
  \bibinfo{author}{\bibfnamefont{S.~R.} \bibnamefont{Jackson}},
  \bibnamefont{and} \bibinfo{author}{\bibfnamefont{F.~W.}
  \bibnamefont{Strauch}}, \bibinfo{journal}{Phys.\ Rev.\ A}
  \textbf{\bibinfo{volume}{81}}, \bibinfo{pages}{012306}
  (\bibinfo{year}{2010}).

\bibitem[{\citenamefont{Gambetta et~al.}(2011)\citenamefont{Gambetta, Motzoi,
  Merkel, and Wilhelm}}]{Gambetta_PRA11}
\bibinfo{author}{\bibfnamefont{J.~M.} \bibnamefont{Gambetta}},
  \bibinfo{author}{\bibfnamefont{F.}~\bibnamefont{Motzoi}},
  \bibinfo{author}{\bibfnamefont{S.~T.} \bibnamefont{Merkel}},
  \bibnamefont{and} \bibinfo{author}{\bibfnamefont{F.~K.}
  \bibnamefont{Wilhelm}}, \bibinfo{journal}{Phys.\ Rev.\ A}
  \textbf{\bibinfo{volume}{83}}, \bibinfo{pages}{012308}
  (\bibinfo{year}{2011}).

\bibitem[{\citenamefont{Motzoi and Wilhelm}(2013)}]{Motzoi_PRA13}
\bibinfo{author}{\bibfnamefont{F.}~\bibnamefont{Motzoi}} \bibnamefont{and}
  \bibinfo{author}{\bibfnamefont{F.~K.} \bibnamefont{Wilhelm}},
  \bibinfo{journal}{Phys.\ Rev.\ A} \textbf{\bibinfo{volume}{88}},
  \bibinfo{pages}{062318} (\bibinfo{year}{2013}).

\bibitem[{\citenamefont{Schutjens et~al.}(2013)\citenamefont{Schutjens, Dagga,
  Egger, and Wilhelm}}]{Schutjens_PRA13}
\bibinfo{author}{\bibfnamefont{R.}~\bibnamefont{Schutjens}},
  \bibinfo{author}{\bibfnamefont{F.~A.} \bibnamefont{Dagga}},
  \bibinfo{author}{\bibfnamefont{D.~J.} \bibnamefont{Egger}}, \bibnamefont{and}
  \bibinfo{author}{\bibfnamefont{F.~K.} \bibnamefont{Wilhelm}},
  \bibinfo{journal}{Phys.\ Rev.\ A} \textbf{\bibinfo{volume}{88}},
  \bibinfo{pages}{052330} (\bibinfo{year}{2013}).

\bibitem[{\citenamefont{Koch et~al.}(2007)\citenamefont{Koch, Yu, Gambetta,
  Houck, Schuster, Majer, Blais, Devoret, Girvin, and Schoelkopf}}]{Koch_PRA07}
\bibinfo{author}{\bibfnamefont{J.}~\bibnamefont{Koch}},
  \bibinfo{author}{\bibfnamefont{T.~M.} \bibnamefont{Yu}},
  \bibinfo{author}{\bibfnamefont{J.}~\bibnamefont{Gambetta}},
  \bibinfo{author}{\bibfnamefont{A.~A.} \bibnamefont{Houck}},
  \bibinfo{author}{\bibfnamefont{D.~I.} \bibnamefont{Schuster}},
  \bibinfo{author}{\bibfnamefont{J.}~\bibnamefont{Majer}},
  \bibinfo{author}{\bibfnamefont{A.}~\bibnamefont{Blais}},
  \bibinfo{author}{\bibfnamefont{M.~H.} \bibnamefont{Devoret}},
  \bibinfo{author}{\bibfnamefont{S.~M.} \bibnamefont{Girvin}},
  \bibnamefont{and} \bibinfo{author}{\bibfnamefont{R.~J.}
  \bibnamefont{Schoelkopf}}, \bibinfo{journal}{Phys.\ Rev.\ A}
  \textbf{\bibinfo{volume}{76}}, \bibinfo{pages}{042319}
  (\bibinfo{year}{2007}).

\bibitem[{\citenamefont{Berry}(2009)}]{Berry_JPA09}
\bibinfo{author}{\bibfnamefont{M.~V.} \bibnamefont{Berry}},
  \bibinfo{journal}{J.~Phys.~A} \textbf{\bibinfo{volume}{42}},
  \bibinfo{pages}{365303} (\bibinfo{year}{2009}).

\bibitem[{\citenamefont{Li et~al.}(2011)\citenamefont{Li, Wu, and
  Wang}}]{Li_PRA11}
\bibinfo{author}{\bibfnamefont{Y.}~\bibnamefont{Li}},
  \bibinfo{author}{\bibfnamefont{L.-A.} \bibnamefont{Wu}}, \bibnamefont{and}
  \bibinfo{author}{\bibfnamefont{Z.~D.} \bibnamefont{Wang}},
  \bibinfo{journal}{Phys.\ Rev.\ A} \textbf{\bibinfo{volume}{83}},
  \bibinfo{pages}{043804} (\bibinfo{year}{2011}).

\bibitem[{\citenamefont{Fasihi et~al.}(2012)\citenamefont{Fasihi, Wan, and
  Nakahara}}]{Fasihi_JPSJ12}
\bibinfo{author}{\bibfnamefont{M.-A.} \bibnamefont{Fasihi}},
  \bibinfo{author}{\bibfnamefont{Y.}~\bibnamefont{Wan}}, \bibnamefont{and}
  \bibinfo{author}{\bibfnamefont{M.}~\bibnamefont{Nakahara}},
  \bibinfo{journal}{J.\ Phys.\ Soc.\ Jpn.} \textbf{\bibinfo{volume}{81}},
  \bibinfo{pages}{024007} (\bibinfo{year}{2012}).

\bibitem[{\citenamefont{Jing et~al.}(2013)\citenamefont{Jing, Wu, Sarandy, and
  Muga}}]{Jing_PRA13}
\bibinfo{author}{\bibfnamefont{J.}~\bibnamefont{Jing}},
  \bibinfo{author}{\bibfnamefont{L.-A.} \bibnamefont{Wu}},
  \bibinfo{author}{\bibfnamefont{M.~S.} \bibnamefont{Sarandy}},
  \bibnamefont{and} \bibinfo{author}{\bibfnamefont{J.~G.} \bibnamefont{Muga}},
  \bibinfo{journal}{Phys.\ Rev.\ A} \textbf{\bibinfo{volume}{88}},
  \bibinfo{pages}{053422} (\bibinfo{year}{2013}).

\bibitem[{\citenamefont{Vacanti et~al.}(2014)\citenamefont{Vacanti, Fazio,
  Montangero, Palma, Paternostro6, and Vedral}}]{Vacanti_NJP14}
\bibinfo{author}{\bibfnamefont{G.}~\bibnamefont{Vacanti}},
  \bibinfo{author}{\bibfnamefont{R.}~\bibnamefont{Fazio}},
  \bibinfo{author}{\bibfnamefont{S.}~\bibnamefont{Montangero}},
  \bibinfo{author}{\bibfnamefont{G.~M.} \bibnamefont{Palma}},
  \bibinfo{author}{\bibfnamefont{M.}~\bibnamefont{Paternostro6}},
  \bibnamefont{and} \bibinfo{author}{\bibfnamefont{V.}~\bibnamefont{Vedral}},
  \bibinfo{journal}{New\ J.\ Phys.} \textbf{\bibinfo{volume}{16}},
  \bibinfo{pages}{053017} (\bibinfo{year}{2014}).

\bibitem[{\citenamefont{Barnes and {Das Sarma}}(2012)}]{Barnes_PRL12}
\bibinfo{author}{\bibfnamefont{E.}~\bibnamefont{Barnes}} \bibnamefont{and}
  \bibinfo{author}{\bibfnamefont{S.}~\bibnamefont{{Das Sarma}}},
  \bibinfo{journal}{Phys.\ Rev.\ Lett.} \textbf{\bibinfo{volume}{109}},
  \bibinfo{pages}{060401} (\bibinfo{year}{2012}).

\bibitem[{\citenamefont{Barnes}(2013)}]{Barnes_PRA13}
\bibinfo{author}{\bibfnamefont{E.}~\bibnamefont{Barnes}},
  \bibinfo{journal}{Phys.\ Rev.\ A} \textbf{\bibinfo{volume}{88}},
  \bibinfo{pages}{013818} (\bibinfo{year}{2013}).

\bibitem{foot2} See Supplementary Information for technical details and full level diagram.

\bibitem{foot1} The generalized $CNOT$ is related to the usual $CNOT$ (which has $\phi_\mu{=}0$ for all $\mu$) by single-qubit $z$ rotations and a global phase.

\bibitem[{\citenamefont{Pedersen et~al.}(2007)\citenamefont{Pedersen, Molmer,
  and Moller}}]{Pedersen_PLA07}
\bibinfo{author}{\bibfnamefont{L.~H.} \bibnamefont{Pedersen}},
  \bibinfo{author}{\bibfnamefont{K.}~\bibnamefont{Molmer}}, \bibnamefont{and}
  \bibinfo{author}{\bibfnamefont{N.~M.} \bibnamefont{Moller}},
  \bibinfo{journal}{Phys.\ Lett.\ A} \textbf{\bibinfo{volume}{367}},
  \bibinfo{pages}{47} (\bibinfo{year}{2007}).

\bibitem[{\citenamefont{Fowler et~al.}(2012)\citenamefont{Fowler, Mariantoni,
  Martinis, and Cleland}}]{Fowler_PRA12}
\bibinfo{author}{\bibfnamefont{A.~G.} \bibnamefont{Fowler}},
  \bibinfo{author}{\bibfnamefont{M.}~\bibnamefont{Mariantoni}},
  \bibinfo{author}{\bibfnamefont{J.~M.} \bibnamefont{Martinis}},
  \bibnamefont{and} \bibinfo{author}{\bibfnamefont{A.~N.}
  \bibnamefont{Cleland}}, \bibinfo{journal}{Phys.\ Rev.\ A}
  \textbf{\bibinfo{volume}{86}}, \bibinfo{pages}{032324}
  (\bibinfo{year}{2012}).

\bibitem[{\citenamefont{Raussendorf et~al.}(2003)\citenamefont{Raussendorf,
  Browne, and Briegel}}]{Raussendorf_PRA03}
\bibinfo{author}{\bibfnamefont{R.}~\bibnamefont{Raussendorf}},
  \bibinfo{author}{\bibfnamefont{D.~E.} \bibnamefont{Browne}},
  \bibnamefont{and} \bibinfo{author}{\bibfnamefont{H.~J.}
  \bibnamefont{Briegel}}, \bibinfo{journal}{Phys.\ Rev.\ A}
  \textbf{\bibinfo{volume}{68}}, \bibinfo{pages}{022312}
  (\bibinfo{year}{2003}).

\bibitem[{\citenamefont{Rosen and Zener}(1932)}]{Rosen_PR32}
\bibinfo{author}{\bibfnamefont{N.}~\bibnamefont{Rosen}} \bibnamefont{and}
  \bibinfo{author}{\bibfnamefont{C.}~\bibnamefont{Zener}},
  \bibinfo{journal}{Phys. Rev.} \textbf{\bibinfo{volume}{40}},
  \bibinfo{pages}{502} (\bibinfo{year}{1932}).

\bibitem[{\citenamefont{Economou et~al.}(2006)\citenamefont{Economou, Sham, Wu,
  and Steel}}]{Economou_PRB06}
\bibinfo{author}{\bibfnamefont{S.~E.} \bibnamefont{Economou}},
  \bibinfo{author}{\bibfnamefont{L.~J.} \bibnamefont{Sham}},
  \bibinfo{author}{\bibfnamefont{Y.}~\bibnamefont{Wu}}, \bibnamefont{and}
  \bibinfo{author}{\bibfnamefont{D.~G.} \bibnamefont{Steel}},
  \bibinfo{journal}{Phys. Rev. B} \textbf{\bibinfo{volume}{74}},
  \bibinfo{pages}{205415} (\bibinfo{year}{2006}).

\bibitem[{\citenamefont{Economou}(2012)}]{Economou_PRB12}
\bibinfo{author}{\bibfnamefont{S.~E.} \bibnamefont{Economou}},
  \bibinfo{journal}{Phys.\ Rev.\ B} \textbf{\bibinfo{volume}{85}},
  \bibinfo{pages}{241401(R)} (\bibinfo{year}{2012}).


\bibitem{foot3} $\widetilde{CZ}$ can be obtained from $CZ$ by applying subsequent single-qubit $z$ rotations and including a global phase.



\end{thebibliography}
\end{document}